\documentclass[prb]{revtex4}
\usepackage{graphicx}
\begin{document}
\begin{abstract}
We consider the Coulomb drag between two metal rings in which the
long range Coulomb interaction leads to the formation of a Wigner
crystal. The first ring is threaded by an Ahranov Bohm flux
creating a persistent current $J_0$.  The second ring is brought
in close proximity to the second and due to the Coulomb
interaction between the two rings a drag current $J_D$ is
produced in the second.  We investigate this system at zero
temperature for perfect rings as well as the effects of
impurities.  We show that the Wigner crystal state can in
principle lead to a higher ratio of drag current to drive current
$J_D/J_0$ than in weakly interacting electron systems.
\end{abstract}
\pacs{PACS numbers: } \vspace{.5in}
\title{Coulomb drag Between One Dimensional Wigner Crystal Rings}
\author{John Baker and A. G. Rojo }
\affiliation{Department of Physics, University of Michigan, Ann
Arbor, MI 48109-1120} \maketitle

\section{Introduction}
Coulomb Drag between current carrying systems has been studied in
numerous papers both experimentally and
theoretically\cite{prl:gramila,prb:zheng,jphysc:rojo,sps:pogre,physica:price}.
The earliest experiments were done in two dimensional systems.
In this configurations the electrons are combined to
two-dimensional layers.  In one such layer a current is driven
and another conducting layer is brought to within close proximity
of the first.  The Coulomb interaction between the electrons in
these two layers causes a transfer of momentum and the second
layer acquires a current.  This current is referred to as the
drag current.

Recently, the phenomenon of Coulomb drag in one dimensional
systems has attracted much attention, particularly with regards
to nanowires and nanotechnology~
\cite{jphyscm:gurevich,prl:raichev,prb:raichev,jetpl:gurevich,physicaE:debray}.
The basic description of the one dimensional case is identical to
that in two dimensions. In all of these systems the time averaged
charge density of the systems are translationally invariant.  The
coupling of the two systems is due to charge fluctuations in the
two systems.

It is known however that the long range character of the Coulomb
interaction can lead to the formation of a Wigner crystal in
which the electrons become localized and form a periodic
lattice.  Numerical simulations on one dimensional systems
indicate that an arbitrarily weak long range interaction will
lead to the formation of a Wigner
crystal\cite{prl:schulz,prb:fano}. This long range nature of the
Coulomb force should be apparent for low electron densities.

In this paper we consider the drag between two such Wigner
crystals. It is evident that this system has a non-uniform charge
distributions and we will show that this in principle can lead to
a much higher value of the drag current.

\section{Classical Drag}
It is the aim of this section to analyze from a classical perspective some aspects of
current drag that we may expect in a quantum mechanical treatment of the Wigner crystal.
 The Wigner crystal is as a system where the electrons are effectively localized, so it
is
reasonable to expect that some aspects of the current drag between Wigner crystals may
be exhibited in a classical system.

\begin{figure}[h]
\vspace{-5pt}
 \hspace{1.cm}
\includegraphics{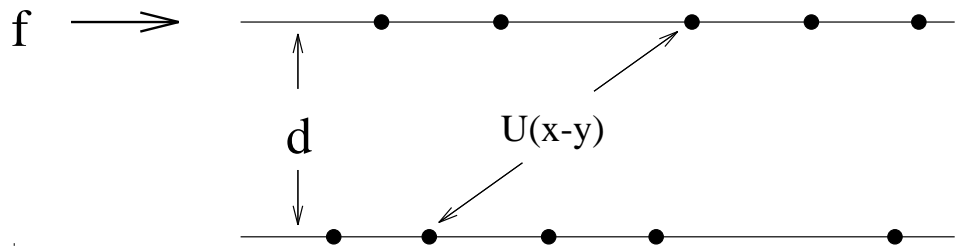}
\vspace{5pt}
\caption{ Illustration of the
current drag setup for considered in this classical analysis.  A
constant force $f$ drives system one.  $U(x-y)$ is the interwire
interaction and $d$ is the distance of separation. }
\label{fig:2wires}
    \end{figure}

 Here, we discuss the current drag between two one dimensional wires consisting
of classical particles interacting via an unscreened Coulomb potential.  The two systems
are close  enough that they
are coupled by the Coulomb interaction between the wires.  This system is illustrated in
Figure~(\ref{fig:2wires}).
\vspace{.1in}
To get an idea of what to expect from such a system we first
consider a "toy model" consisting of two lines of equally spaced
particles interacting via a periodic potential and moving under a
viscous drag force.  The main difference between this system and
that in Figure~(\ref{fig:2wires}) is that the spacing between
particles is fixed so that the internal dynamics of the system are
ignored.  The description of such a system has been considered in
an analysis by Cladis, Parks and Daniels\cite{prl:cladis} in the
context of the vortex dynamics in type II superconductors.

In this analysis, two rigid chains of equally spaced particles are
taken in the presence of a viscous drag force $F_{d}=\alpha
\dot{x}$.  Since each chain is rigid, its position can be denoted
by the position of one of its particles.  Call $x_{1}$ the
position of chain $1$ and $x_{2}$ the position of chain $2$. The
interaction between the two chains is then periodic in $x_1-x_2$
and modeled by
$F_{i}=A\sin[(2\pi/a)(x_{1}-x{2})]$\cite{note:periodic} where $a$
is the lattice spacing.  In addition, the first chain is driven by
a constant force $f$. In steady state the equations of motion of
the two chains are given by

\[ \alpha\dot{x_{1}}=-A\sin[(2\pi/a)(x_{1}-x_{2})] + f
\]

\[ \alpha\dot{x_{2}}=A\sin[(2\pi/a)(x_{1}-x_{2})]
\]
    Setting $\phi=(x_{1}-x_{2})$,  results in the following equation for the
difference between the velocities of the two chains.
\[\dot{\phi}=f/\alpha-\frac{2A}{\alpha}\sin(2\pi\phi/a)
\]
with solution
\[T_{0}=\int_{0}^{a/2}\frac{d\phi}{\frac{f}{\alpha}-\frac{2A}{\alpha}\sin(2\pi\phi/a)}
\]
where $T_{0}$ is the time for the two chains to slip one complete cycle (distance
$a/2$).

Carrying out the integral results in
\[ T_{0}=\left\{ \begin{array}{ll}
    \infty & \mbox{ for $f<2A/ \alpha$ (locking) } \\
    \frac{a}{2}\frac{1}{(f/\alpha)^{2}-(2A/\alpha)^{2}}
        & \mbox{ for $f \geq 2A/\alpha$ (slipping)}
    \end{array}
    \right. .
\]
This simple analysis shows the interesting result that for suitable conditions of the
parameters $f$ and $A$, the two chains will be locked together with the drag current
equal to the drive current.

We now wish to investigate the behavior of this classical system when internal dynamics
are included.   We take therefore as a model of the dynamics of these classical
particles the following Fokker-Planck equation
\begin{equation}
\label{eq:FP}
\frac{\partial}{\partial t} P(\vec{x},\vec{y},t)=\sum_{i}
D\left(\frac{\partial^{2}}
{\partial x_{i}^{2}} + \frac{1}{k_{B}T}\frac{\partial}{\partial x_{i}}
F_{i}\right)
P(\vec{x},\vec{y},t)
+\sum_{j} D\left(\frac{\partial^{2}}{\partial y_{j}^{2}} +\frac{1}{k_{B}T}
\frac{\partial}{\partial y_{j}} F_{j}\right) P(\vec{x},\vec{y},t)
\end{equation}
Here, $\vec{x}=(x_{1},x_{2}...,x_{N})$ labels the positions of $N$ particles in
wire $1$, $\vec{y}=(y_{1},y_{2}...,y_{N})$  labels the positions of particles in
wire $2$, $D$ is the diffusion constant and $T$ is the temperature.
$P(\vec{x},\vec{y},t)$ is then the probability
of having the $N$ particles in layer $1$ at positions $\vec{x}$ and the $N$ particles
in
layer $2$ at position $\vec{y}$ at time $t$.  The forces present are
\[F_{i}=f + \frac{\partial}{\partial x_{i}}V(\vec{x}) + \frac{\partial}
{\partial x_{i}}U(\vec{x},\vec{y})
\]

 in wire $1$ and
 \[F_{j} =\frac{\partial}{\partial y_{j}}V(\vec{y}) + \frac{\partial}
{\partial y_{j}}U(\vec{x},\vec{y})
\]
in wire $2$. The interaction $U$ is taken to be the
unscreened Coulomb interaction between the wires, $V$ is the unscreened Coulomb
interaction within each wire and $f$ is a constant force driving the particles in
wire $1$.  The basic problem is to determine the drag current
produced in wire $2$ due to the moving charges in wire $1$.

\begin{figure}[!]
\resizebox{4in}{!}{\includegraphics{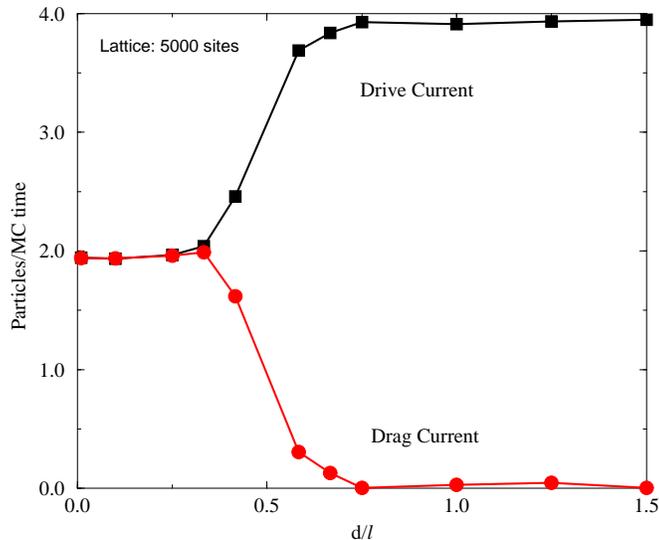}} \vspace{5pt}
\caption{ Plot of drag current against wire separation  for two
one-dimensional wires. } \label{fig:1Dplots}
\end{figure}

In the numerical analysis of this problem we performed Monte Carlo
simulations on discrete systems.  The conducting systems are
partitioned into a lattice with the particles occupying positions
at the lattice points.  The dynamics of the simulation are
determined by equation~(\ref{eq:FP}) which is approximated
numerically as follows. For each particle, one of two possible
random directions is chosen .  If the particle lowers its energy
by moving one lattice site in that direction, then the move is
accepted. Otherwise the move is accepted with probability
$e^{-\beta \Delta E}$, where $\beta = 1/k_{B}T$ and $\Delta E$ is
the change in the particles energy upon making this move.  This
process is completed iteratively for each particle until a steady
state condition is reached.  Simulations were performed to
determine the dependence of the drag current on separation between
the two systems.  The results are shown in
Figure~(\ref{fig:1Dplots}) \vspace{.1in}

The plot of Drag Current vs. Separation shows the fact that the sum of the drive
and drag currents is a constant,
$\overline{J}_{d}+\overline{J}_{0}=\frac{NDf}{k_{B}T}$ as can be seen by averaging
the right hand side of equation~(\ref{eq:FP}), ($ RHS=
\sum_{i}\frac{\partial}{\partial x_{i}}J_{0} + \sum_{j}\frac{\partial}{\partial
y_{j}}J_{d}$).   It also shows that the drag current is significant only when the
distance of separation is on
the order of the interparticle spacing or smaller.

The main feature of interest in this plot, however, is that the
drag current, $J_d$, is essentially equal in magnitude to the
drive current, $J_0$, at small interwire seperations.  As the
separation between the wires is increased the drag current shows
a smooth decrease in magnitude.  The main difference between this
plot and the toy model (where internal dynamics are ignored) is
that there does not appear to be an abrupt locking transition.
However at sufficiently large values of the interaction parameter,
$A$ the two chains are essentially locked together with the drag
current equal in magnitude to the drive current. We will now
examine to what extent this classical effect is exhibited in a
quantum mechanical system.

\section{Wigner crystal Ring}
\subsection{Disorder Free Case}
In this section we calculate the non-dissipative drag between two
perfect Wigner crystal rings at zero temperature.  The basics of
non-dissipative drag effects has been considered in previous
papers~\cite{prl:rojo_mahan,prb:baker}.  The basic system of
interest here consists of two metallic rings one of which is
threaded by an Ahranov-Bohm flux, $\Phi$.  This flux shifts the
ground state, $E_g$, of the ring to a current carrying state which
is calculated according to
\[
J=\frac{\partial E_g}{\partial \Phi}.
\]
   If another ring is brought close to the first one, then the Coulomb interaction will
cause a drag current to be produced in the second ring.  Since
this is the ground state of the two ring system this current is
non-dissipative and persists as long as the magnetic flux is
present in the first ring.

In these works the two rings have uniform charge densities and the
drag effect is due to coupling of charge fluctuations in each
ring.  In a system with zero disorder, the relative magnitude of
the drag current produced is typically a small fraction of the
driving current.  For two such systems separated by a distance
$\sim 200\AA$ the drag current is on the order of $\sim 10^{-4}$
times the current in the driving system.~\cite{prl:rojo_mahan}
Based on the analysis of the last section, we expect that the drag
current of two Wigner crystals could in principle be as large as
the driving current.

\begin{figure}[!]
\resizebox{4in}{!}{\includegraphics{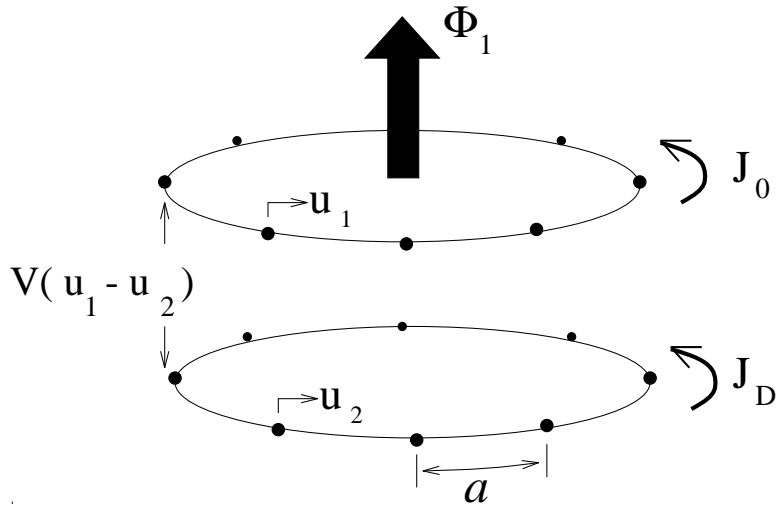}}
\vspace{5pt} \caption{ Schematic illustration of the two ring
Wigner crystal setup. The crystal period is $a$ and $u_1(x,t)$
denotes the local displacement of ring $1$ and $u_2(x,t)$ denotes
the local displacement of ring $2$. Ring $1$ has an Arhanov-Bohm
flux,$\Phi_1$, present which creates a persistent current $J_0$ in
ring $1$.  The interaction between the two crystals, $V(u_1-u_2)$
is a function of the relative displacement of the two cryatals and
creates a drag current $J_D$ in ring $2$. } \label{fig:two_rings}
\end{figure}

The model of the one dimensional Wigner crystal that we use here
is that proposed in references (\cite{prb:glazman}) and
(\cite{prb:krive}).  In this model the Wigner crystal is viewed as
an elastic chain of spinless electrons.  In the continuum limit,
the Wigner crystal of length $L$, in the presence of an Ahranov
Bohm flux, $\Phi$, is described by the Lagrangian~\cite{prb:krive}

\begin{equation}
\label{eq:lagrange1ring} \mathcal{L}=\int
dx\left[\frac{m}{2a}\left(\dot{u}^2-s^2(u')^2\right)+\frac{\hbar}{L}
\frac{\Phi}{\Phi_0}\dot{u}\right]
\end{equation}

where $a$ is the crystal period, m is the electron mass, s is the velocity of sound in
the crystal and $\Phi_0$ is the flux quantum.  The field variable $u(x,t)$ describes the
local displacement of the chain at a point $x$ at time $t$.  If the field variable is
expanded
in a fourier series
\[
u=\frac{1}{\sqrt{L}}\sum_k c_k e^{ik\theta}
\]
we get,
\[
\mathcal{L}=(\mu/2)\sum_k\left(\dot{c}_k^2-s^2k^2c_k^2\right)-\frac{\hbar}{L}\tilde{\Phi}\dot{c}_0
, \] where $\tilde{\Phi}=\Phi/\Phi_0$.
 It is clear that the flux will only couple to the zero mode of the ring.  So in dealing
with the persistent current of a disorder free Wigner crystal ring, the internal
dynamics of the ring may be ignored and only the rotation of the ring as a whole is
relevant. We wish to study the interaction between two such Wigner crystal rings as
shown in Figure~(\ref{fig:two_rings}).
\vspace{.1in}
We take as a model for two disorder free interacting Wigner crystal rings the following
Hamiltonian.
\begin{equation}
\label{eq:WC_ringL}
H=\frac{1}{2M}\left(-i\hbar\frac{\partial}{\partial
u_1}-\frac{e}{c}\Phi_1\right)^2 +
\frac{1}{2M}\left(-i\hbar\frac{\partial}{\partial
u_2}-\frac{e}{c}\Phi_2\right)^2 + V_0\cos\frac{2\pi}{a}(u_1-u_2)
\end{equation}
where $M=Nm$ is the total mass of each ring, $V_0$ represents the
strength of the inter-ring interaction, $\Phi_1$ is the flux
through ring 1, $\Phi_2$ is the flux through ring 2 and $u_{1,2}$
represents the displacement of ring $1,2$ respectively. The first
two terms in Equation~(\ref{eq:WC_ringL}) represent the
individual dynamics of rings $1$ and $2$ respectively.  The third
term represents the interaction between the two rings. It is
evident that the interaction between the chains will be periodic
in the relative displacement between the two crystals.  For
simplicity we model this interaction as a cosine term, however
the basic features of this analysis should not depend on the
particular form of the periodic potential used.

Transforming to center of mass and relative angular coordinates
$\phi=\frac{2 \pi}{L}(u_1-u_2)$ and
$\theta=\frac{\pi}{L}(u_1+u_2)$, we have in dimensionless form
\begin{equation}
\label{eq:WC_ringH}
\left[\left(i\frac{\partial}{\partial \phi}+ \frac{\alpha}{2}\right)^2 +
\frac{1}{4}\left(i\frac{\partial}{\partial \phi} + \beta\right)^2+
q\cos(N\phi)\right]\Psi=\epsilon\Psi
\end{equation}
where $\alpha=\Phi_1-\Phi_2$, $\beta=\Phi_1+\Phi_2$, $q=V_0(MR^2/\hbar^2)$ and
$\epsilon=E(MR^2/\hbar^2)$.  Our task is now to solve equation~(\ref{eq:WC_ringH}) for
the eigenenergy $\epsilon$.

  The hamiltonian, equation~(\ref{eq:WC_ringH}), separates into two parts
$H=H_\phi+H_\theta$ with energies
$\epsilon=\epsilon_\phi+\epsilon_\theta$.  The ground state
wavefunction is a product
 $\Psi(\phi,\theta)=S(\theta)T(\phi)$.
  The equation for the
center of mass coordinates is readily solved giving
 \[
 S(\theta)=\frac{1}{\sqrt{2\pi}}e^{in'\theta}
 \]
 \begin{equation}
\label{eq:COMH}
 \epsilon_\theta=\frac{1}{4}(n'+\beta)^2 ;n'=0,\pm1,\pm2,...
 \end{equation}
 The equation for the relative coordinate wavefunction is
 \begin{equation}
\label{eq:relativeH} \left[-\left(i\frac{\partial}{\partial
\phi}+ \frac{\alpha}{2}\right)^2
-q\cos(N\phi)+\epsilon_\phi\right]T(\phi)=0
\end{equation}
 For $\alpha=0$ this reduces to Mathieu's equation.   The energy eigenvalues
correspond to the periodic solutions $T(0)=T(2\pi)$.  We therefore
substitute the Fourier expansion
 \[
 T=\sum_{-\infty}^\infty c_ne^{in\phi}
 \]
 into equation~(\ref{eq:relativeH}) and obtain the following recursion relation for the coefficients
$c_n$
 \[
 \frac{q}{2}c_{n+1} + (\epsilon-[n-\alpha]^2)c_n +
 \frac{q}{2}c_{n-1}=0.
 \]   The energy eigenvalues may
be found numerically to any desired accuracy by truncating the set
of equations for some $c_N$ and searching for the value of
$\epsilon$ which gives a zero value for the appropriate
determinant of coefficients.~\cite{book:jordan}

\begin{figure}[!]
\vspace{-5pt} \hspace{.5cm}
\resizebox{3in}{!}{\includegraphics{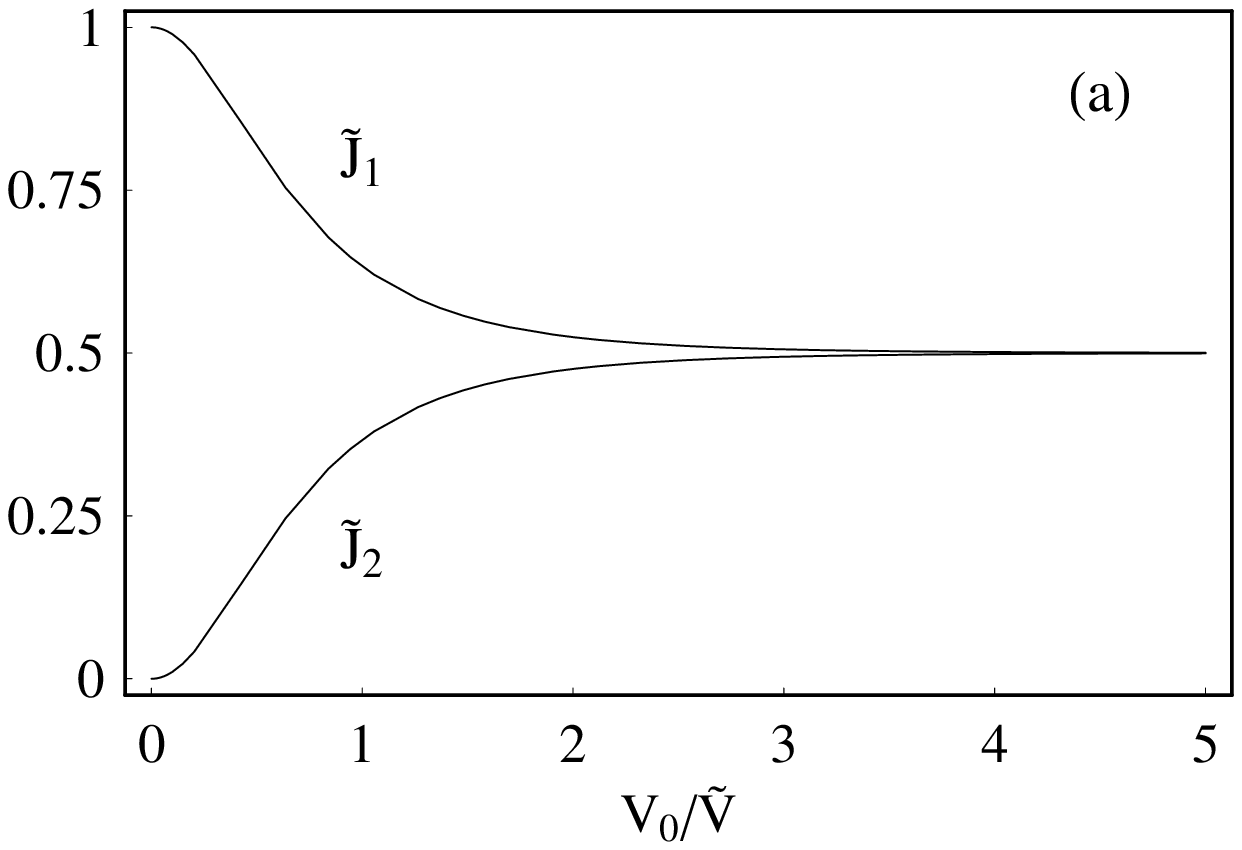}}
\resizebox{3in}{!}{\includegraphics{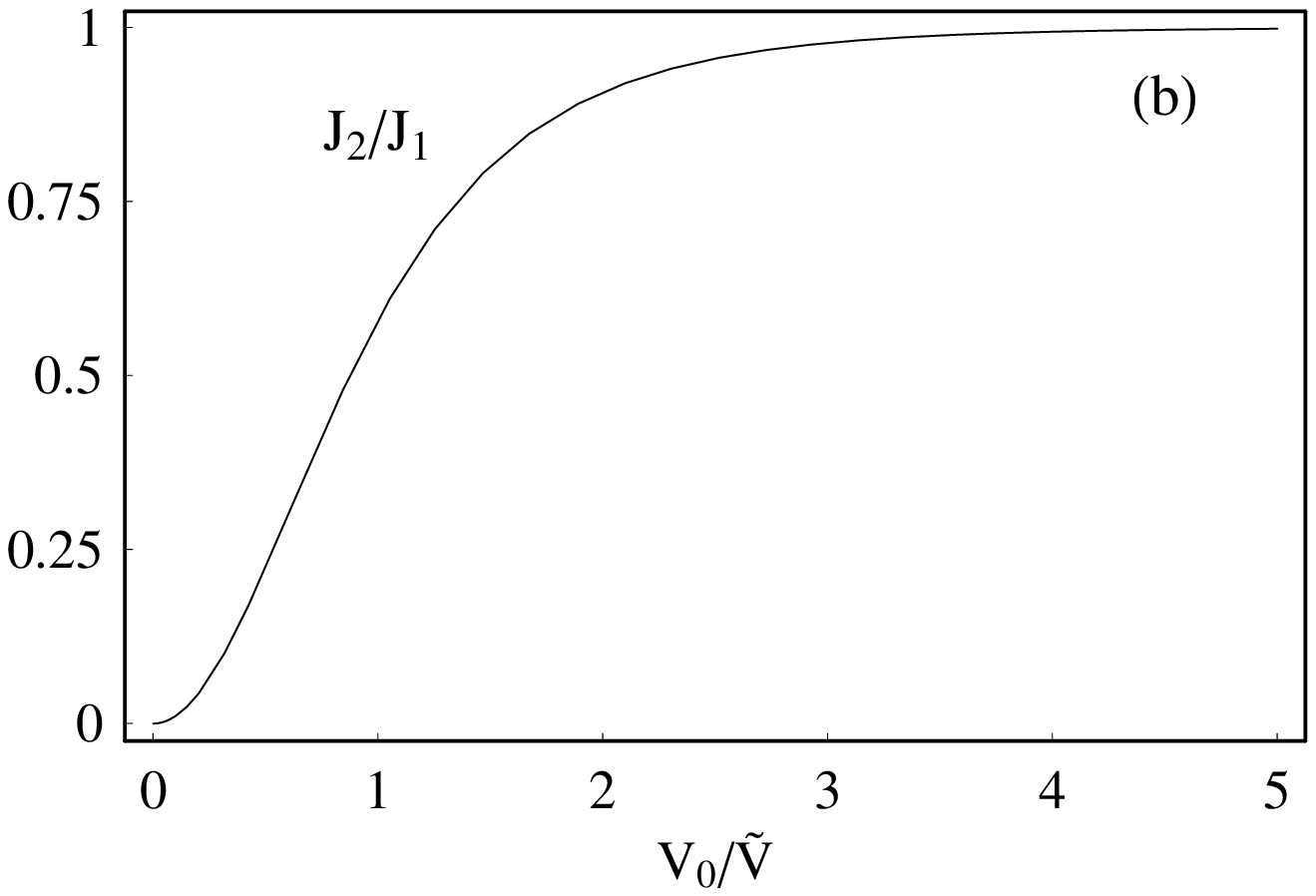}} \vspace{5pt}
\caption{ Plots of drive and drag current against the interaction
strength. $V_0$ is shown in units of
$\tilde{V}=\frac{N\hbar^2}{mR^2}$ and the current
$\tilde{J}=J/J_1(0)$ where $J_1(0)$ denotes the current in ring
$1$ at $V_0 =0 $.}
 \label{fig:wigner_Jvsq}
\end{figure}

 The energy eigenvalues determined in this way, combined with the results of
equation~(\ref{eq:COMH})
allow the currents in each ring to be determined from $J_1=\frac{\partial
\epsilon}{\partial \Phi_1}$ and
 $J_2=\frac{\partial \epsilon}{\partial \Phi_2}|_{\Phi_2=0}$.\cite{prb:baker}. The
results are shown in Figure~(\ref{fig:wigner_Jvsq}).

 \vspace{.1in}

Figure~(\ref{fig:wigner_Jvsq}) shows that the drag current does
approach the drive current in magnitude as the interaction $V_0$
is increased which is qualitatively similar to the classical
result. For large values of the interaction strength,
$q/N\gtrsim3$ or
\begin{equation}
\label{eq:Lock_cond}
 V_0\gtrsim3\hbar^2 N/mR^2,
 \end{equation}
  the two Wigner crystals have
essentially the same current with each having half the current
that one ring would have if isolated.

\begin{figure}[!h]
\vspace{-5pt} \hspace{.5cm}
\resizebox{4in}{!}{\includegraphics{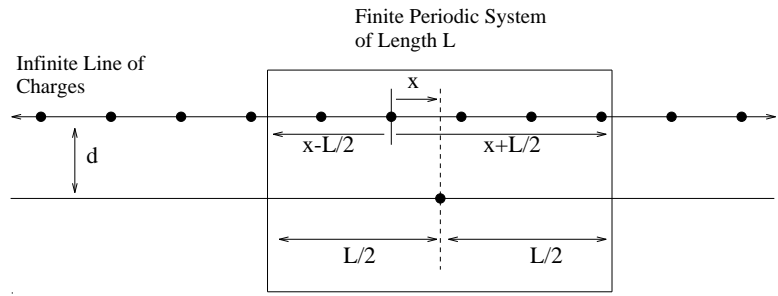}}
\vspace{5pt} \caption{ Schematic model of a periodic system of
electrons.  The potential on an electron centered at $x$ and a
distance $d$ from an infinite line of electrons is obtained by
taking a finite segment of length $L$ of the infinite chain
centered at x.}
 \label{fig:amplitude_approx}
\end{figure}

 In order to relate the interaction in equation~(\ref{eq:WC_ringL}) to the true Coulomb interaction
 we approximate the amplitude $V_0$ as follows.
 We want to calculate the potential felt by an electron a distance $d$ from a
 finite periodic system of length $L$ and period $a$.  We model our system by taking a length $L$
  of an infinite line of electrons centered at $x$.  This system is shown schematically in
Figure~(\ref{fig:amplitude_approx}).

\vspace{.1in}

This potential energy is given by
\[
V=\sum_{n=-\frac{1}{a}(L/2-x)}^{\frac{1}{a}(L/2+x)}
\frac{e^2}{\sqrt{(x-na)^2+d^2}}
\]
 which can be rewritten as
\[
V=\int_{-\frac{1}{a}(L/2-x)}^{\frac{1}{a}(L/2+x)} dz
\frac{e^2}{\sqrt{(x-az)^2+d^2}}\sum_{n=-\infty}^{\infty}\delta(z-n)
\]
\[
= \int_{-\frac{1}{a}(L/2-x)}^{\frac{1}{a}(L/2+x)} dz
\frac{e^2}{\sqrt{(x-az)^2+d^2}} \sum_{l=-\infty}^{\infty}e^{i
2\pi lz}
\]
\[
=-\sum_{l=-\infty}^{\infty}\int_{-L/2}^{L/2} \frac{dz'}{a}
\frac{e^2}{\sqrt{z'^2+d^2}} e^{i 2\pi \frac{l}{a}(x-z)}
\]
\[
=\int_{-L/2}^{L/2}\frac{dz'}{a} \frac{e^2}{\sqrt{z'^2+d^2}}  + 2
\sum_{l=1}^{\infty}\left[\int_{-L/2}^{L/2} \frac{dz'}{a}
\frac{e^2}{\sqrt{z'^2+d^2}} \cos\left(\frac{2\pi
l}{a}z'\right)\right]\cos\left(\frac{2\pi l}{a} x\right)
\]

The first term is a constant in $x$ therefore we ignore it.  For
small separation distances $d/a \lesssim 1$, higher $l$ terms may
be ignored and we keep only the $l=1$ term. This yields finally
\[
V \approx \frac{4}{a}\left[\int_{0}^{L/2} dz'
\frac{\cos\left(\frac{2\pi}{a}z'\right)}{\sqrt{z'^2+d^2}}\right]
\cos\left(\frac{2\pi}{a} x\right)
\]
  We can now identify the amplitude $V_0$ in
Equation~(\ref{eq:WC_ringL})for a crystal with N electrons as
\begin{equation}
\label{eq:WC_amplitude}
 V_0=N \frac{4}{a}\int^{L/2}_{0}
dz' \frac{\cos\frac{2\pi}{a}z'}{\sqrt{z'^2+d^2}}.
 \end{equation}

\[
=N \frac{4}{a}\int^{L/2a}_{0} dz \frac{\cos(2 \pi z)
}{\sqrt{z^2+d^2}}.
\]
As mentioned in the introduction for a one dimensional system, a
long range interaction will lead to the formation of a Wigner
crystal\cite{prl:schulz,prb:fano}. The long range nature of the
Coulomb force should be apparent for electron densities, $n
\lesssim \frac{1}{10}a_B$ where $a_B$ is the Bohr radius for the
particular ring material.\cite{prl:ivanov} For a typical
mesoscopic ring radius of $100 \mu m$, crystal lattice constant
$a=10a_B$ and $a_B \approx 2 \AA$ we have $L/a >10^5$ so the upper
limit in the integral in Equation~(\ref{eq:WC_amplitude}) may be
taken to infinity which yields
\[
V_0=\frac{4Ne^2}{a}K_0(2\pi\frac{d}{a})
 \]
where $K_0$ is the zeroth order Bessel function. This implies an
exponential decrease in the amplitude of the interaction with
increasing separation between the two Wigner crystals.  Comparison
with Equation~(\ref{eq:Lock_cond}) shows that the two crystals
will be "locked" together at a distance $d/a \approx 4$.  So two
very low density perfect Wigner crystals will be essentially
locked together when separated by a distance of four times the
lattice constant or less.

\subsection{Impurity Effects}

We now consider the effects of impurities on the dynamics of the
Wigner crystal. In the presence of an impurity, the phase of the
Wigner crystal will be pinned.  In this case the crystal ring
cannot rotate as a whole so the current as presented in the last
subsection is not possible.  It was shown, however, by Rice et
al.~\cite{prl:rice}, that a new type of current is possible,
namely the tunneling of solitons through the impurity barrier. It
is the tunneling of solitons that we will investigate in this
section.

\begin{figure}[!]
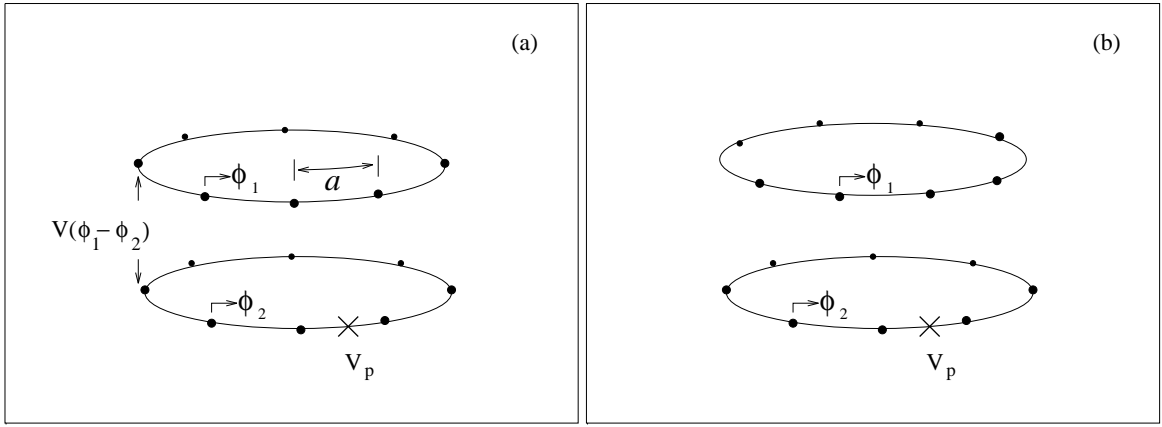

\hspace{.5cm} \vspace{-5pt}
\resizebox{3in}{!}{\includegraphics{rings_phase0b.eps}}
\resizebox{3in}{!}{\includegraphics{rings_phasePib.eps}}
\vspace{10pt} \caption{ Schematic illustration of the two ring
Wigner crystal setup. The phase $\phi_{1,2}=\frac{2\pi}{a}
u_{1,2}(x,t)$ where $u_{1,2}$ is the local displacement of the
Wigner crystal. The interaction between the two rings is $V(\phi_1
- \phi_2) = W[1-\cos\left(\phi_1-\phi_2\right)]$. Ring $2$ has an
impurity of strength $V_p$ that couples to the phase $\phi_2$.
Figure~(\ref{fig:two_ringsPhase}b) shows a phase shift in the
relative coordinate $(\phi_1-\phi_2)$ of $\pi$ relative to
Figure~(\ref{fig:two_ringsPhase}a). } \label{fig:two_ringsPhase}
\end{figure}

We are interested in knowing the ratio of the drag current in ring
two, to the drive current in ring one.  We therefore take as an
assumption the presence of a current in ring one and look for the
relative magnitude of the drag current in ring two in the presence
of a single impurity.  This system is shown schematically in
Figure~(\ref{fig:two_ringsPhase}). \vspace{.1in}
Using Equation~(\ref{eq:lagrange1ring}), we take as a description of our two ring Wigner
crystal system, the following Lagrangian
\begin{equation}
\label{eq:lagrangeWC}
 \mathcal{L}= \mathcal{L}_1+\mathcal{L}_2-W\left[1-\cos(\phi_1-\phi_2)\right]
 \end{equation}
 Where
 \begin{equation}
\label{eq:lagrangeRing1}
 \mathcal{L}_1=\frac{ma}{8\pi^2}\left\{\frac{\partial\phi_1}{\partial t}^2 -
s^2\frac{\partial\phi_1}{\partial x}^2\right\},
 \end{equation}
 \begin{equation}
\label{eq:lagrangeRing2}
 \mathcal{L}_2=\frac{ma}{8\pi^2}\left\{\left(\frac{\partial\phi_2}{\partial t}\right)^2 -
s^2\left(\frac{\partial\phi_2}{\partial x}\right)^2\right \}
-V_p\delta(x)\left[1-\cos(\phi_2)\right],
 \end{equation}
and the phase $\phi_{1,2}=u_{1,2}(x,t)(2\pi/a)$.

The first two terms in Equation~(\ref{eq:lagrangeWC}) are the
Lagrangians describing the dynamics of rings $1$ and $2$
respectively.  The last term describes the interaction between
the two rings.  We follow the work of Krive et
al.~\cite{prb:krive} and introduce the dimensionless parameter
\[
\alpha=\frac{\pi\hbar}{msa}
\]
which characterizes the magnitude of the quantum fluctuations of the Wigner crystal, and
the parameter
\[
T_s=\hbar s
\] which characterizes the energy scale in a Wigner crystal of length $L$. We will be
concerned with stiff crystals that are weakly fluctuating characterized by $\alpha \ll
1$.

The basic approach that we will use here is a semiclassical
approximation used by Larkin and Lee~\cite{prb:larkin} in the
context of charge density waves and elaborated by
Kleinert\cite{book:kleinert} for a single particle tunneling
through a barrier . In this approach the probability amplitude of
a soliton tunneling through an impurity is given by $e^{A/\hbar}$
where $A$ is the action associated with the tunneling trajectory
obtained by minimizing the function $\int dt \int dx
\mathcal{L}(\tau)$, where $\mathcal{L}(\tau)$ is the Lagrangian,
Equation~(\ref{eq:lagrangeWC}), with $t$ replaced by the imaginary
time $\tau=it$. This approach amounts to approximating the path
integral of the propagator by its value at the classical
trajectory.\cite{book:kleinert}

We now want to answer the following question: given a current in
ring $1$ that causes a rotation of the phase by $2\pi$, what is
the probability that the phase of ring $2$ will also rotate by
$2\pi$? It is apparent that there are three possiblities:

$i$) $\phi_1 \rightarrow 2\pi, \phi_2 \rightarrow 0$: the phase in ring $2$ will remain
pinned and ring 1 will rotate on top of the relative potential between the two ending
with a relative phase difference $(\phi_1-\phi_2)=2\pi$.

$ii$) $\phi_1 \rightarrow 2\pi, \phi_2 \rightarrow 2\pi$:  the phase of ring $2$ will
tunnel through the impurity thus increasing by $2\pi$ and ending in phase with $\phi_1$.

$iii$) $\phi_1 \rightarrow 2\pi, \phi_2 \rightarrow -2\pi$: the phase of ring $2$ will
counter-rotate with that in ring $1$ ending with a relative phase difference,
$(\phi_1-\phi_2)=4\pi$.

Here we have limited our consideration to single tunneling
trajectories.  In other words, each shift of the phase of each
ring occurs due to a single tunneling event and ignore all
multistep tunneling trajectories. Thus we only consider strong
coupling between rings, $WL/(\frac{N \hbar^2}{mR^2})\gtrsim 1$,
and strong pinning $\alpha V_p \gg T_s$.  We now calculate the
tunneling amplitudes for each case above.

In case ($i$), the Lagrangian in Equation~(\ref{eq:lagrangeWC}) reduces to
 \begin{equation}
\label{eq:lagrangeCase1}
 \mathcal{L}=-\frac{ma}{8\pi^2}\left\{\left(\frac{\partial\phi_1}{\partial \tau}\right)^2 +
s^2\left(\frac{\partial\phi_1}{\partial x}\right)^2\right\} -
W\left[1-\cos(\phi_1)\right]
 \end{equation}
 This is simply the problem of a single Wigner crystal ring in the presence of a
periodic pinning potential.  The classical wave equation resulting from this Lagrangian
in real time, is
\[
\frac{ma}{8\pi^2}\left\{\ddot{\phi_1}-s^2\stackrel{''}{\phi_1}\right\}+W\sin(\phi_1)=0
\]
This is the sine-Gordon equation and it admits the exact soliton solution
\[
\phi_1=4\tan^{-1}\left[\exp\pm\left(\frac{\tilde{x}-u\tilde{t}}{\sqrt{1-u^2}}\right)\right
]
\]
where $\tilde{x}=\frac{\omega_0}{s} x$, $\tilde{t}=\omega_0 t$ and
\[
 \omega_0=\sqrt{\frac{4\pi^2W}{ma}}
 .\]
The $+$ sign corresponds to a soliton rotating in the clockwise direction, the $-$ sign
corresponds to an antisoliton rotating in the counterclockwise direction and $u$ is the
soliton/antisoliton velocity.  The boundary conditions on such a solution are
$\phi(-L/2,t)=\phi(L/2,t)$.  A solution to this equation that matches this boundary
condition is the soliton/antisoliton solution and has the form
\[
\phi_1(x,t)=4 \tan^{-1}\left[\frac{u
\sinh\left(\tilde{x}/\sqrt{1-u^2}\right)}{\cosh\left(u\tilde{t}/
\sqrt{1-u^2}\right)}\right].
\]  This solution has the following interpretation: at $t=0$ the phase of the crystal
is $0$, as $t$ increases, the phase at $x=0$ locally increases to $2\pi$ creating a
region around $x=0$ with phase $2\pi$.  As $t$ increases this region with $2\pi$ phase
propagates symmetrically from $x=0$ at speed $u$ towards $x=-L/2$ and $x=-L/2$ until the
entire ring is at $\phi=2\pi$.  The Lagrangian, Equation~(\ref{eq:lagrangeCase1}) has
been considered in the context of charge density wave tunneling by Bogachek et
al.~\cite{prb:bogachek} , and there it is noted that the homogeneous (space independent)
soliton solution is advantageous in the action against any spatially dependent solution
for a finite system.  The homogenous solution corresponds to a uniform rotation of the
Wigner crystal as a whole and has the form~\cite{prb:bogachek}
\begin{equation}
\label{eq:solutionCase1} \phi_1(\tau)=4\tan^{-1}
\exp(\omega_0\tau) .
 \end{equation}
 This solution describes the uniform increase of the phase of the entire ring from $0$
to $2\pi$ in a time $\tau_0 \sim 1/\omega_0$.  The action associated with this solution
is by Equation~(\ref{eq:lagrangeCase1})
 \begin{equation}
\label{eq:actionCase1}
A_1=-\frac{8WL}{\omega_0}=-8\left[\frac{WL}{\left(\frac{N}{mR^2}\right)}\right]^{1/2}
 \end{equation}

In case ($ii$), the relative difference between the two phases
does not increase, while the phase of the center of mass of the
two ring system increases by $2\pi$. We assume that during the
tunneling process the two rings remain in phase to minimize the
potential energy between them. If we rewrite
Equation~(\ref{eq:lagrangeWC}) using $\phi=\phi_1-\phi_2$ and
$\theta=\phi_1+\phi_2$, we get the relevant Lagrangian for case
($ii$) to be

\begin{equation}
\label{eq:lagrangeCase2}
\mathcal{L}=-\frac{ma}{8\pi^2}\left\{\left(\frac{\partial\theta}{\partial
\tau}\right)^2 + s^2\left(\frac{\partial\theta}{\partial
x}\right)^2
\right\}-V_p\delta(x)\left[1-\cos\left(\frac{\theta}{2}\right)\right]
\end{equation}

This is the problem of a single Wigner crystal ring tunneling
through an impurity.  We  consider the case of strong crystal
pinning, $\alpha V_p\gg T_s$.

The case of soliton tunneling through an impurity in a strongly
pinned Charge Density wave was first considered by Larkin and
Lee~\cite{prb:larkin}.  In their analysis, the tunneling process
is broken into two stages.  The first stage consists of a rapid
tunneling of a length $l_0$ of the crystal through the impurity in
a short time $t_0$ and leaves the crystal in a distorted state of
high potential energy.  The second stage consists of a slow
relaxation of the crystal from this distorted state back into a
state that minimizes its potential energy.

To describe the first stage of the process Larkin and Lee proposed the following trial
solution to Equation~(\ref{eq:lagrangeCase2})
\[
\theta=\frac{2\pi t}{t_0}\left(1-\frac{|x|}{l_0}\right)
\]
for $0<t<t_0$ and $|x|<l_0$.

It is evident that this solution interpolates between the initial state, $\theta=0$ for
all $x$ at $t=0$, and the final state $\theta=2\pi$ for $x=0$ at $t=2\pi$. The second
stage of the process takes place away from the impurity $|x|>l_0$ and obeys the equation
of motion
\[
\ddot{\theta}+s^2\theta''=0.
\]
  The solution to this stage of the tunneling should obey the
periodic boundary conditions of the ring and match the solution
for the first stage  at $x=-L/2,L/2$. This problem has been
considered in the work of Krive et al~\cite{prb:krive} where the
proposed trajectory for the relaxation stage has the form
\[ \phi=\pi \pm 2\arctan\left(\frac{(\tau-\tau_s)s}{|x|}\right)
.\] The action associated with the tunneling stage of the process
is
\[
A_{2t}=-\frac{\hbar}{\alpha}\left(C_1+C_2\frac{\alpha
V_p}{T_h}\right).
\]
where $C_{1,2}$ are constants.
The action associated with the relaxation stage is
\[
A_{2r}=-\frac{\hbar}{\alpha}\ln\left(\frac{L}{2 l_0}\right).
\]
The length $l_0$ is obtained by minimizing the total action $A_{2T}=A_{2t}+A_{2r}$ with
the result that
\[
l_0=\frac{\hbar s}{C \alpha V_p}
\]
and
\begin{equation}
\label{eq:actionCase2a}
A_{2T}=-\frac{\hbar}{\alpha}\left\{\ln\left(\frac{\alpha V_p}{T_s}\right)+C_3\right\}
\end{equation}
where $C$ is a constant and $C_3$ is a constant assumed to be of
the order unity~\cite{prb:larkin}.
Equation~(\ref{eq:actionCase2a}) shows the interesting result
that the action is only logarithmically dependent on the impurity
for strong pinning.

In case ($iii$) above, the center of mass of the two ring system does not move while the
phase of the relative coordinates increases by $4\pi$.  The relevant Lagrangian for this
system is given by
\begin{equation}
\label{eq:lagrangeCase3}
\mathcal{L}=-\frac{ma}{8\pi^2}\left\{\left(\frac{\partial\phi}{\partial
\tau}\right)^2 + s^2\left(\frac{\partial\phi}{\partial x}\right)^2
\right\}-V_p\delta(x)\left[1-\cos\left(\frac{\phi}{2}\right)\right]
- W\left[1-\cos\left(\frac{\phi}{2}\right)\right].
\end{equation}

This is identical to the lagrangian in Equation~(\ref{eq:lagrangeCase2}) with the
addition of the term proportional to $W$.  Since the probability for this process goes
as $e^A$ it is expected that the probability for this case will be reduced relative to
that in case ($ii$) by $\sim e^{-W}$ and can therefore be neglected.

We are now in a position to analyze the relative magnitude of the drag current $J_D$ to
the drive current $J_0$.  This ratio is given by
\begin{equation}
\label{eq:drag_Drive}
\frac{J_D}{J_0}=\frac{e^{\frac{A_{2T}}{\hbar}}}{e^{\frac{A_1}{\hbar}}+e^{\frac{A_{2T}}{\hbar}}}
\end{equation}

It is clear that if $|A_1| \gg |A_{2T}|$ the two crystal will
essentially be locked together.  For purposes of illustration we
consider a stiff crystal with $\alpha=.1$ and consider values of
$V_p$ for stiff pinning $\alpha V_p\gg T_s$. \vspace{.1in}
\begin{figure}[!]
\vspace{-5pt} \hspace{.5cm}
\resizebox{3in}{!}{\includegraphics{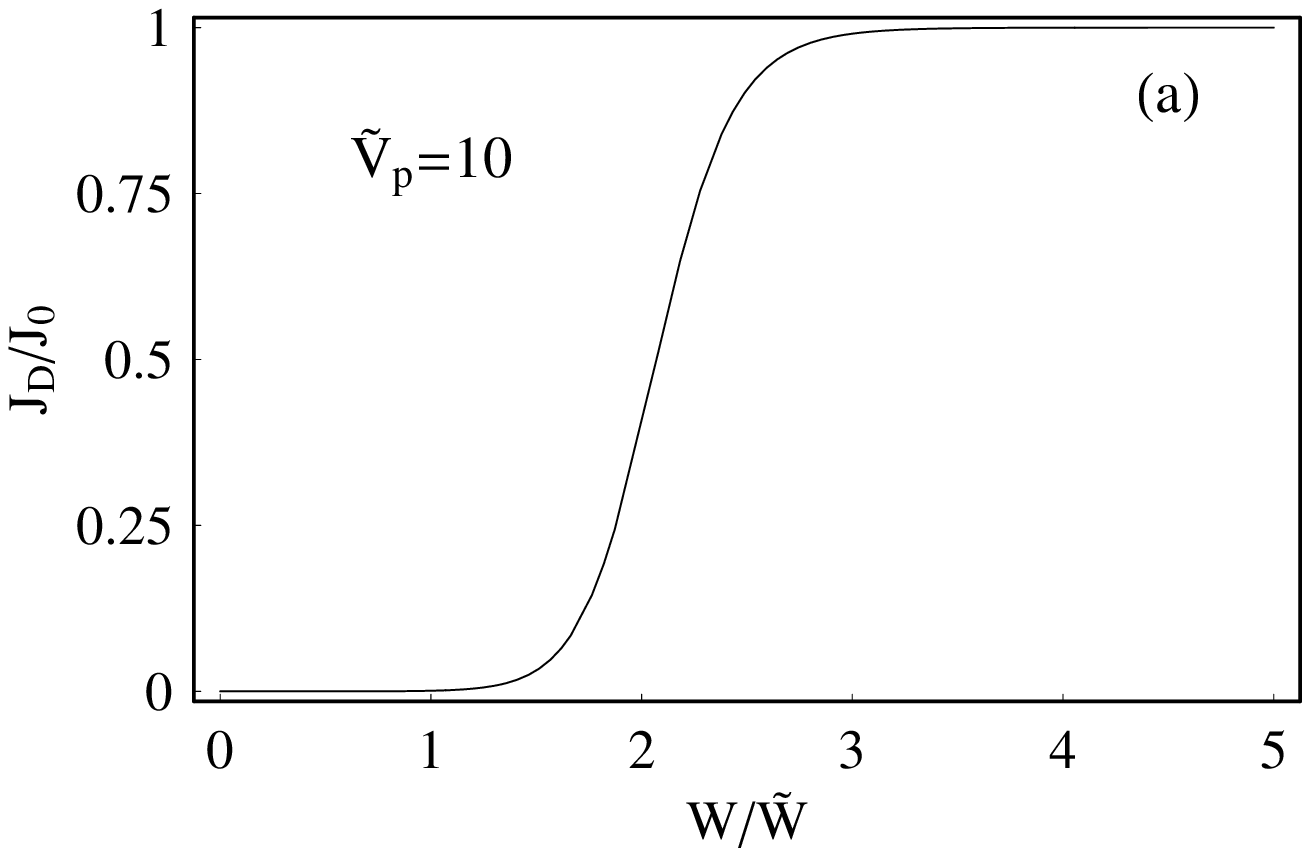}}
\resizebox{3in}{!}{\includegraphics{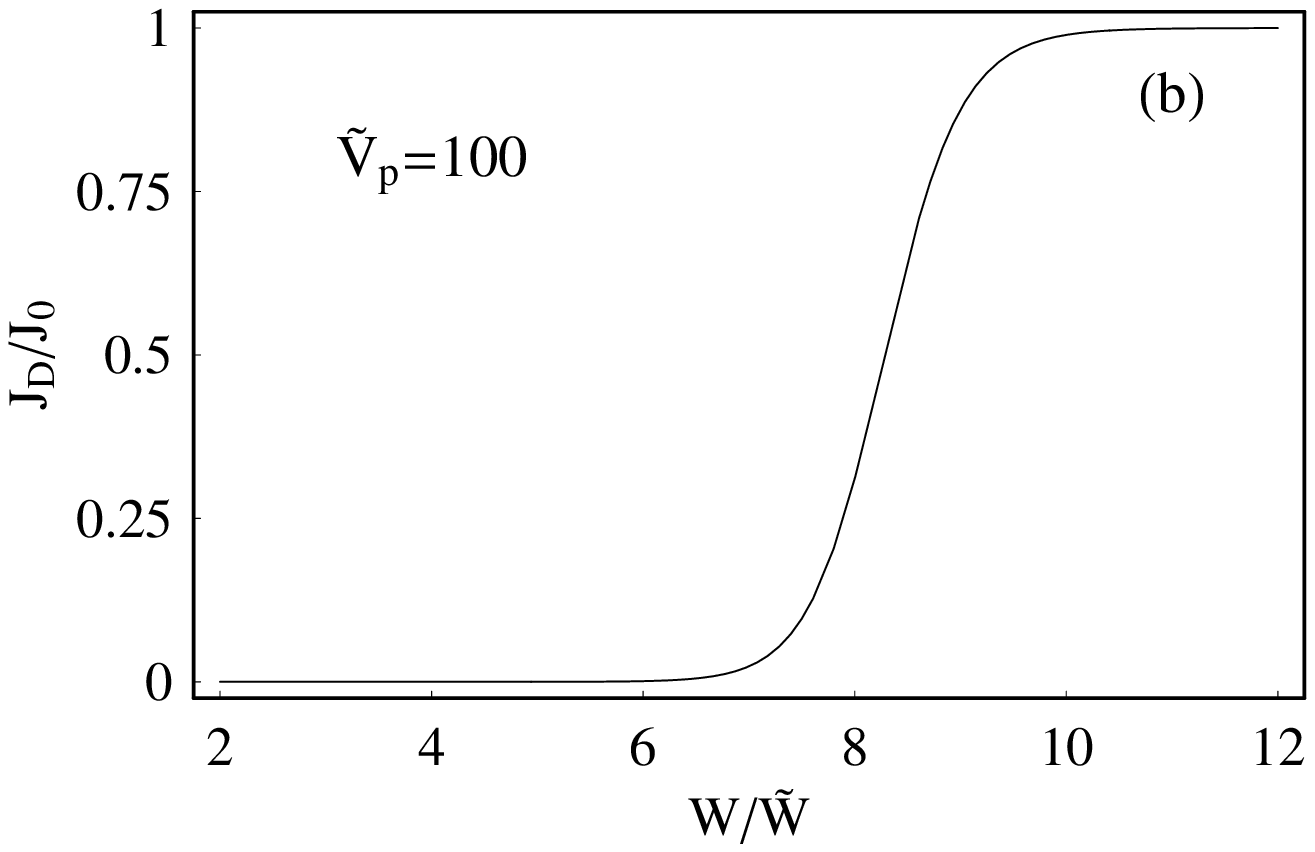}} \vspace{5pt}
\caption{ Plots $J_D/J_0$ versus interaction strength W, for
$\alpha=.1$ and $\alpha V_P/T_s=10$ and $100$. The interaction
strength is shown in units of $\tilde{W}=\frac{N \hbar}{mR^2}$
and $\tilde{V}_p=\frac{\alpha V_p}{T_s}$}
 \label{fig:impurity_lock}
\end{figure}
Figure~(\ref{fig:impurity_lock}) shows that the impurity effects
are not that drastic in determining the value of $W$ where the
locking of the two crystals occur.  This is due to the fact that
the impurity potential enters logarithmically into the tunneling
action.  Comparison with Figure~(\ref{fig:wigner_Jvsq}b) shows
that the distance at which the two Wigner crystals become "locked"
together is essentially the same as in the impurity free case. The
main effect of the impurity is then in the magnitude of the
persistent current.

We can estimate the magnitude of these currents by noting that if
the two currents are locked together, then the magnitude of the
persistent current for a given flux $\Phi$ should be the same as a
single Wigner crystal ring with twice the mass. This current has
been calculated at zero temperature to be~\cite{prb:krive}
\begin{equation}
\label{eq:currentWeakPot} J_0 \sim T_s\left(\frac{T_s}{\alpha
V_p}\right)^{1/\alpha}e^{-C_3/\alpha}\cos\left(2\pi\frac{\Phi}{\Phi_0}\right).
\end{equation}  We see that the current in
Equation~(\ref{eq:currentWeakPot}) is much smaller than the
disorder free persistent current due to the $(1/V_p)^{1/\alpha}$
factor.

In summary we have analyzed the Coulomb drag between two
one-dimensional Wigner crystal rings.  For sufficiently large
interaction between rings the drag current is essentially equal in
magnitude to the drive current. For an impurity free ring this
"locking" of currents occurs at a separation distance of $d
\approx 4a$ with $a$ the crystal period. We analyzed the effect of
a single impurity in the drag ring and found that the two currents
are "locked" at essentially the same distance and the major effect
of the impurity is to significantly decrease the magnitude of the
persistent current.


\begin{thebibliography}{99}
\bibitem{note:periodic} As pointed out in reference\cite{prl:cladis}, the exact form
of the interaction is not important for the results so long as
the force is periodic in the vortex lattice.
\end{thebibliography}
\end{document}